\documentclass{emulateapj}
\usepackage{apjfonts}
\usepackage{amsmath}
\usepackage{multirow}
\usepackage{array}
\usepackage{epsfig}
\usepackage[usenames]{color}
\allowdisplaybreaks[1]
\newcommand{\comment}[1]{}

\definecolor{purple}{RGB}{160,32,240}
\newcommand{\peter}[1]{}

\newcommand{\Msun}{M_{\odot}}
\newcommand{\massname}{characteristic}

\begin{document}

\shortauthors{BEHROOZI, WECHSLER \& CONROY}
\shorttitle{On the Lack of Evolution in Galaxy Star Formation Efficiency}
\submitted{Submitted to ApJ Letters}
\title{On the Lack of Evolution in Galaxy Star Formation Efficiency}

\author{Peter S. Behroozi, Risa H. Wechsler}
\affil{Kavli Institute for Particle Astrophysics and Cosmology; Physics Department, Stanford University; 
Department of Particle Physics and Astrophysics, SLAC National  Accelerator Laboratory; 
Stanford, CA 94305}
\author{Charlie Conroy}
\affil{Department of Astronomy \& Astrophysics, University of California at Santa Cruz, Santa Cruz, CA 95064}

\begin{abstract}
Using reconstructed galaxy star formation histories, we calculate the instantaneous efficiency of galaxy star formation (i.e., the star formation rate divided by the baryon accretion rate) from $z=8$ to the present day.  This efficiency exhibits a clear peak near a \massname{} halo mass of $10^{11.7}\Msun$, which coincides with longstanding theoretical predictions for the mass scale relevant to virial shock heating of accreted gas.  Above the \massname{} halo mass, the efficiency falls off as the mass to the minus four-thirds power; below the \massname{} mass, the efficiency falls off at an average scaling of mass to the two-thirds power.  By comparison, the shape and normalization of the efficiency change very little since $z=4$.  We show that a time-independent star formation efficiency simply explains the shape of the cosmic star formation rate since $z=4$ in terms of dark matter accretion rates.  The rise in the cosmic star formation from early times until $z=2$ is especially sensitive to galaxy formation efficiency. The mass dependence of the efficiency strongly limits where most star formation occurs, with the result that two-thirds of all star formation has occurred inside halos within a factor of three of the \massname{} mass, a range that includes the mass of the Milky Way.
\end{abstract}

\section{Introduction}

\label{intro}

Theorists have long predicted that galaxy formation is most efficient near a mass of $\sim 10^{12}\Msun$ based on analyses of supernova feedback, cooling times, and galaxy number counts \citep{Silk77,Rees77,Dekel86,White78,Blumenthal84}.  More recently, hydrodynamical simulations have indicated that the host dark matter halo mass strongly influences gas accretion onto galaxies \citep{Birnboim03,Keres05,Dekel06}.  For low halo masses, these simulations predict that gas accretes in cold filaments (``cold mode accretion'') directly to the galaxy disk, efficiently forming stars.  Above a transition halo mass of  $10^{11}$ to $10^{11.5}\Msun$ (which is predicted to be redshift-independent for $z<3$), a shock develops at the virial radius which heats accreting gas (``hot mode accretion'') and rapidly quenches star formation \citep{Dekel06}.

To test these predictions, we use previously-generated statistical reconstructions of the galaxy---halo connection for all observable galaxies \citep{BWC12} to compare the average star formation rate in galaxies to the average baryon accretion rate as a function of halo mass and time.  This approach allows us to directly test for a \massname{} mass scale in the efficiency of star formation in halos.  We summarize the reconstruction method in \S \ref{s:methodology}, present our main results in \S \ref{s:results}, and conclude in \S \ref{s:conclusions}.  Throughout this work, we assume a \cite{chabrier-2003-115} initial mass function, the \cite{bc-03} stellar population synthesis model, and the dust model in \cite{blanton-roweis-07}.  We additionally assume a flat, $\Lambda$CDM cosmology with parameters $\Omega_M = 0.27$, $\Omega_\Lambda = 0.73$, $h=0.7$, $n_s = 0.95$, and $\sigma_8 = 0.82$.

\begin{figure*}
\vspace{-7ex}
\hspace{-12ex}\begin{tabular}{cc}
\epsfig{file=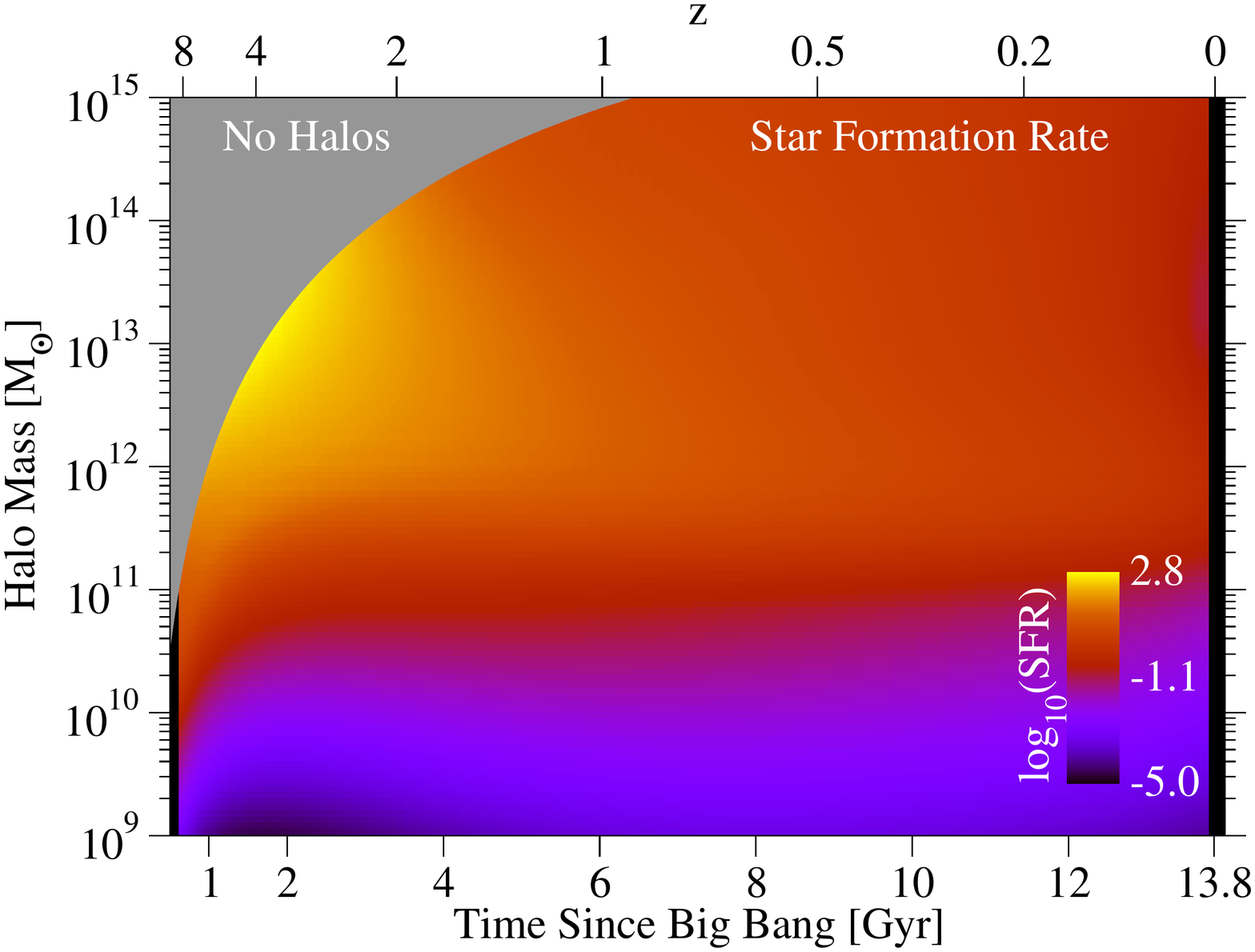,scale=0.33}&\epsfig{file=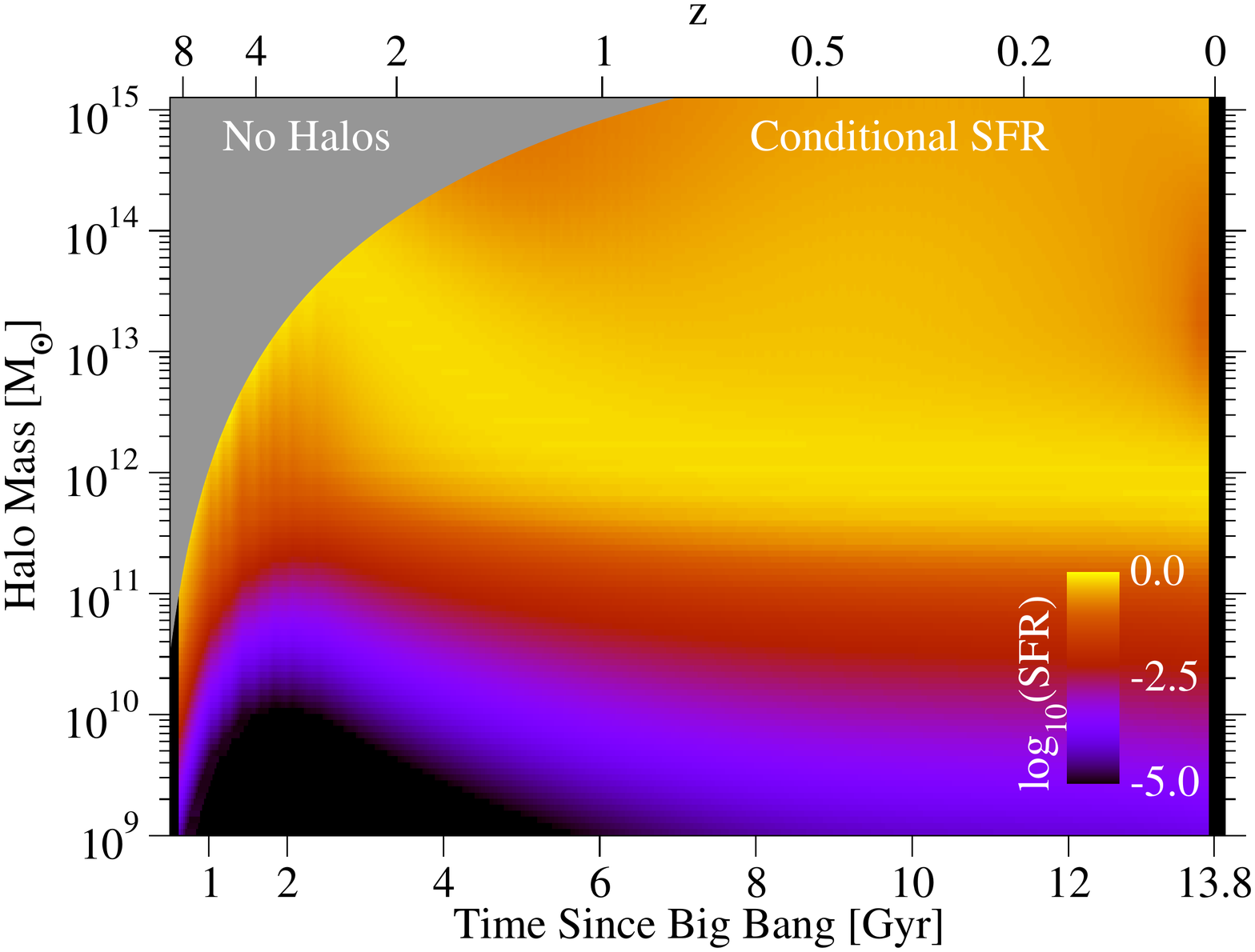,scale=0.33}\\[-5ex]
\epsfig{file=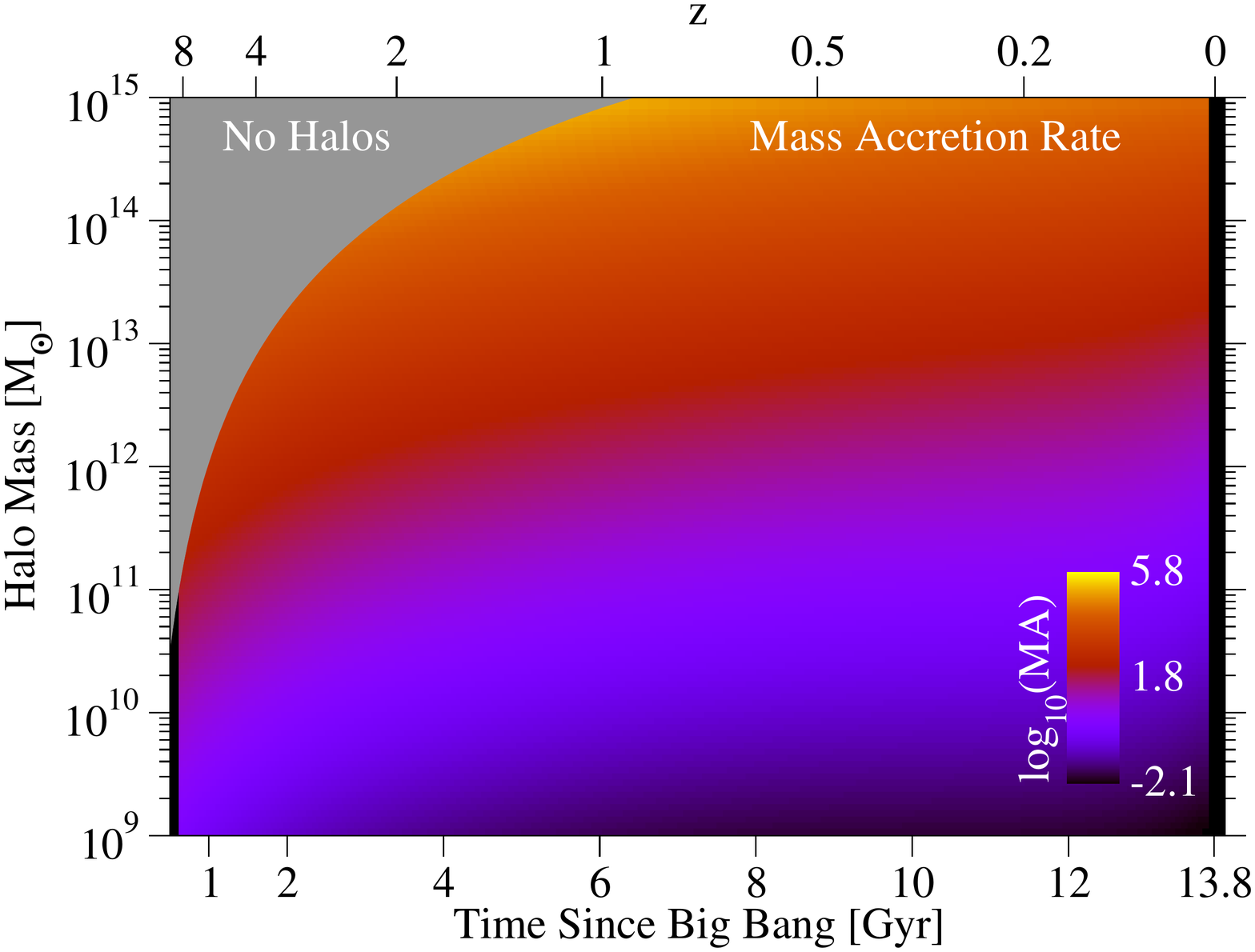,scale=0.33}&\epsfig{file=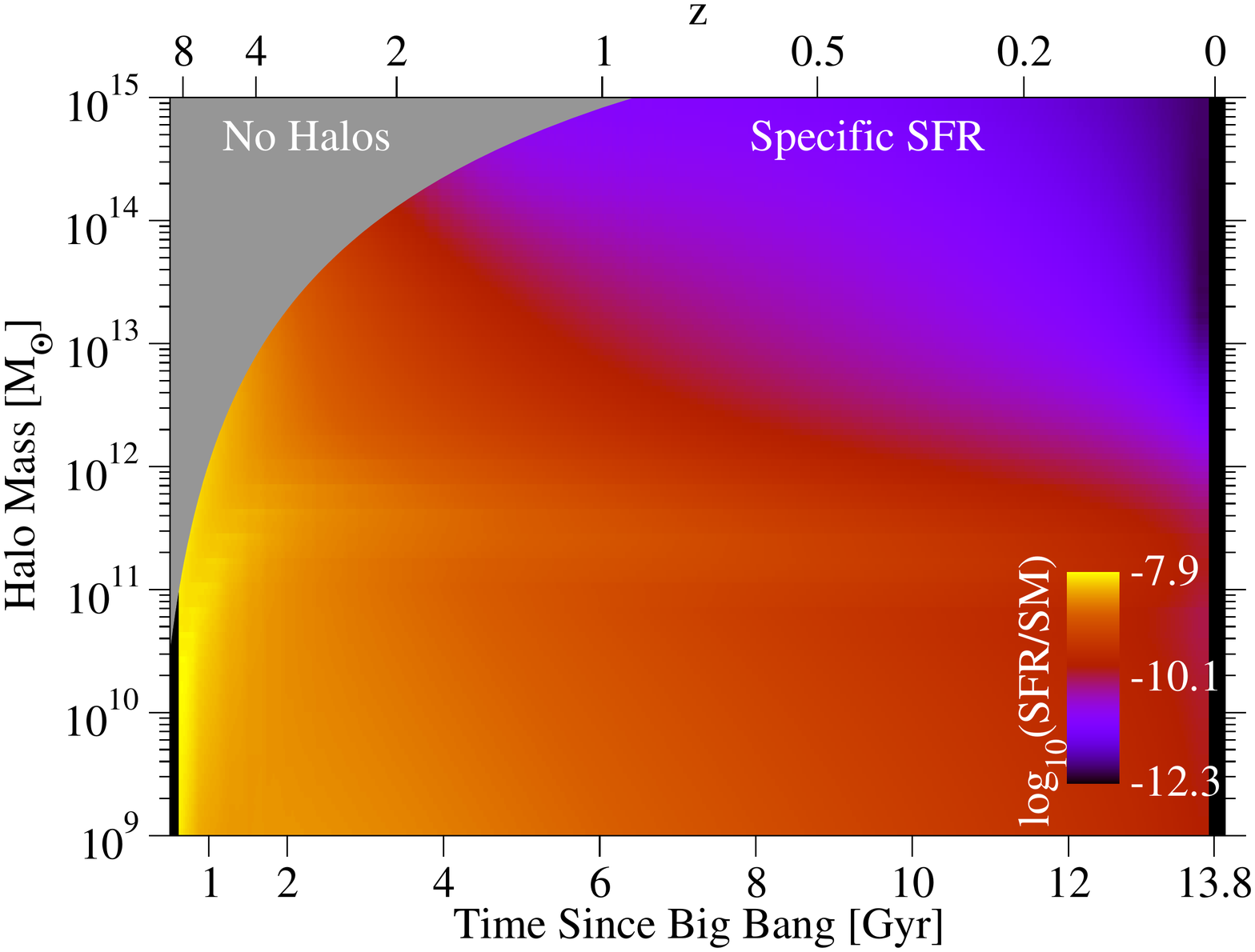,scale=0.33}\\[-5ex]
\end{tabular}
\begin{center}
\vspace{-3ex}
\makebox[1.6\columnwidth][l]{\epsfig{file=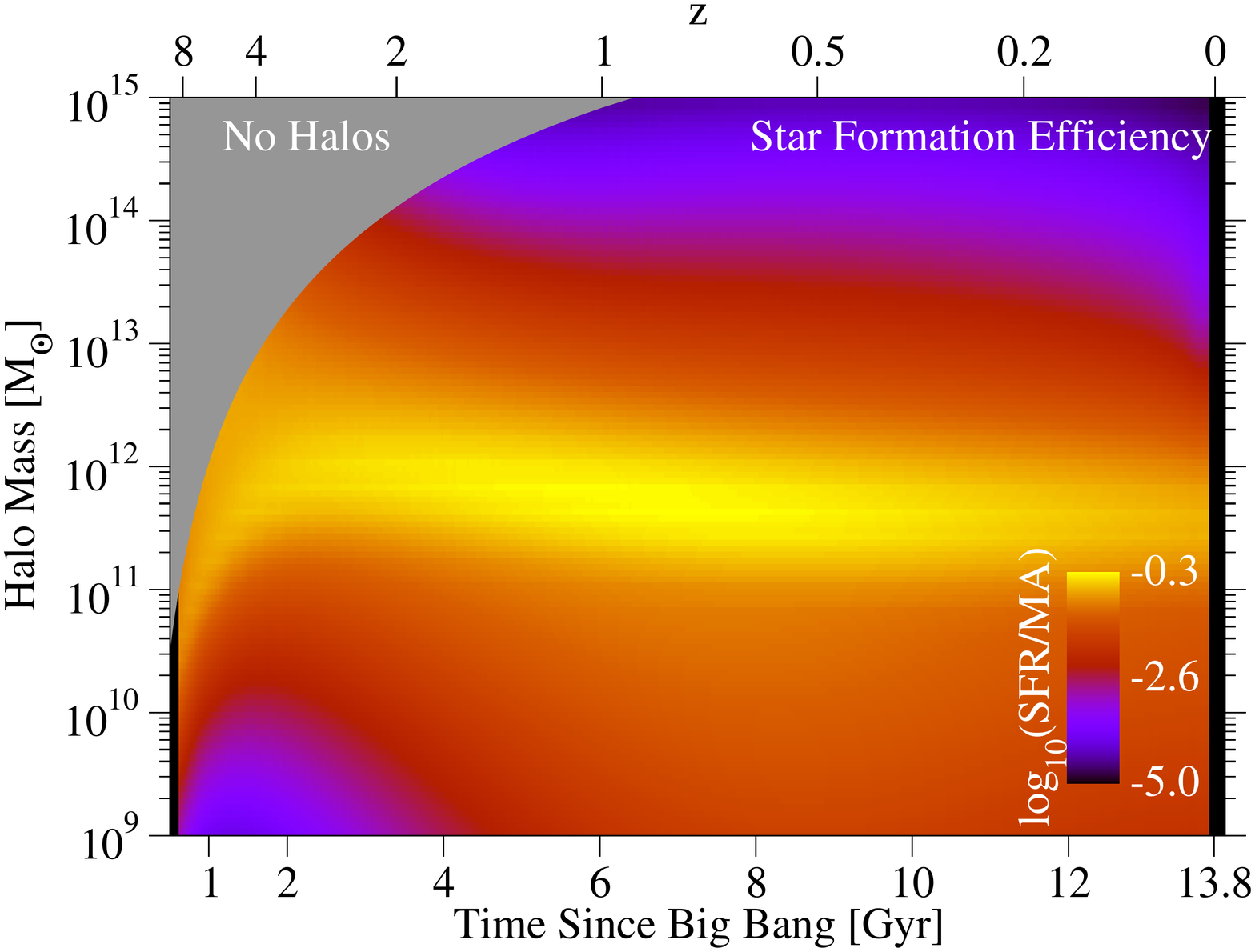,scale=0.4}}
\end{center}
\vspace{-3ex}
\caption{\textbf{Top-left} panel: Star formation rate as a function of halo mass and cosmic time in units of $\Msun$ yr$^{-1}$.  The grey shaded band excludes halos not expected to exist in the observable universe.  \textbf{Top-right} panel: Conditional star formation rate as a function of halo mass and cosmic time, in units of the maximum star formation rate at a given time.  \textbf{Middle-left} panel: baryonic mass accretion rate (MA) in halos as a function of halo mass and time, in units of $\Msun$ yr$^{-1}$.  \textbf{Middle-right} panel: the star formation rate to stellar mass ratio, in units of yr$^{-1}$, as a function of halo mass and time.  There is a roll-off towards higher halo masses; however, the normalization and characteristic mass are strongly redshift-dependent. \textbf{Bottom} panel: instantaneous star formation efficiency  (star formation rate divided by baryonic mass accretion rate) as a function of halo mass and time.}
\label{f:bya}
\end{figure*}

\vspace{5ex}

\section{Statistical Reconstructions}

\label{s:methodology}

To summarize our reconstruction technique (fully detailed in \citealt{BWC12}), we link galaxies observed at different redshifts to halos in a dark matter simulation using an extremely flexible parametrization for the stellar mass--halo mass relation over cosmic time $(SM(M,z))$.\footnote{Six parameters control the relation at $z=0$ (a characteristic stellar mass, a characteristic halo mass, a faint-end slope, a massive-end shape, a transition region shape, and the scatter in stellar mass at fixed halo mass); for each of these parameters, two more variables control the evolution to intermediate ($z\sim1$) and high ($z>3$) redshifts.}  Any choice of $SM(M,z)$, applied to halo merger trees, will result in predictions for the galaxy stellar mass function, average specific star formation rates of galaxies, and the cosmic star formation rate.  We use a Markov Chain Monte Carlo (MCMC) method to constrain $SM(M,z)$ to match observations of these quantities from $z=8$ to $z=0$.  We calculate uncertainties from a wide range of statistical and systematic effects (including uncertainties from stellar population synthesis models, dust models, stellar population history models, the faint-end slope of the stellar mass function, scatter between stellar mass and halo mass, etc.; see \citealt{Behroozi10,BWC12}), mitigating potential biases from, e.g., limited observational constraints at high redshifts.   Alternate initial mass functions are not modeled; these would primarily cause uniform normalization shifts in stellar masses and star formation rates, which would not affect our conclusions.  We use free priors on the functional form of $SM(M,z)$, but we require non-negative star formation rates in all galaxies, and we require that the stellar mass to halo mass ratio is always less than the cosmic baryon fraction.

We combine observational constraints from over $40$ recent papers (see \citealt{BWC12} for a full list).  These include results from SDSS and from PRIMUS \citep{Moustakas12}, which self-consistently recover stellar mass functions from $z=1$ to $z=0$ over a wide area of the sky.  At high redshifts, we include recent measurements of stellar masses and star formation rates to $z=8$ \citep{Bouwens11b,Bouwens11,McLure11,BORG12}.  Notably, measurements of the cosmic star formation rate now agree with the evolution of the stellar mass density \citep{Bernardi10,Moster12,BWC12}.  

For simulation data (halo mass functions, merger rates, and accretion histories), we extensively use the \textit{Bolshoi} simulation \citep{Bolshoi}.  This dark matter simulation follows 2048$^3$ particles in a periodic, comoving volume 250 $h^{-1}$ Mpc on a side using the \textsc{art} code \citep{kravtsov_etal:97,kravtsov_klypin:99}; it has a mass resolution of $1.9 \times 10^8$ $\Msun$ and a force resolution of 1 $h^{-1}$ kpc.  The adopted flat, $\Lambda$CDM cosmology (see \S\ref{intro}) is consistent with the latest WMAP7+BAO+H$_0$ results \citep{wmap7}.  Simulation analysis was performed using the \textsc{rockstar} halo finder \citep{Rockstar} and merger tree code in \cite{BehrooziTree}.

\section{Results}

\label{s:results}

\subsection{Strong Mass Dependence for the Star Formation Efficiency}

\begin{figure}
\hspace{-5ex}\epsfig{file=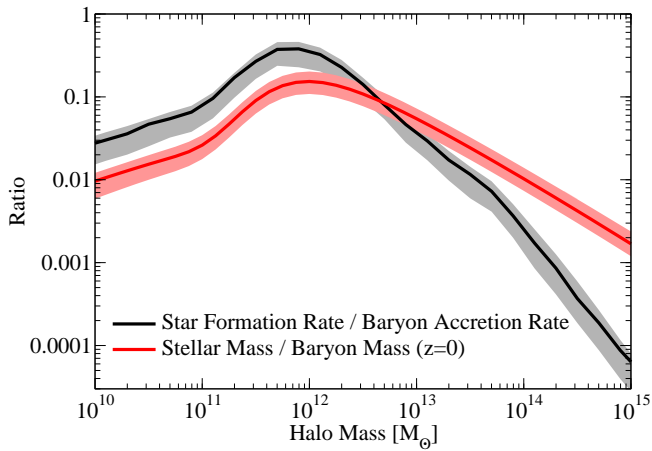,scale=0.33}
\caption{The instantaneous star formation efficiency compared to the integrated star formation efficiency at $z=0$ (i.e., stellar mass over the product of the baryon fraction with the halo mass).  The shaded bands around each line show the one-sigma uncertainty contours.  The integrated efficiency has a different peak and profile, as discussed in the text.}
\label{f:integrated}
\end{figure}

\begin{figure*}
\hspace{-12ex}\begin{tabular}{cc}
\epsfig{file=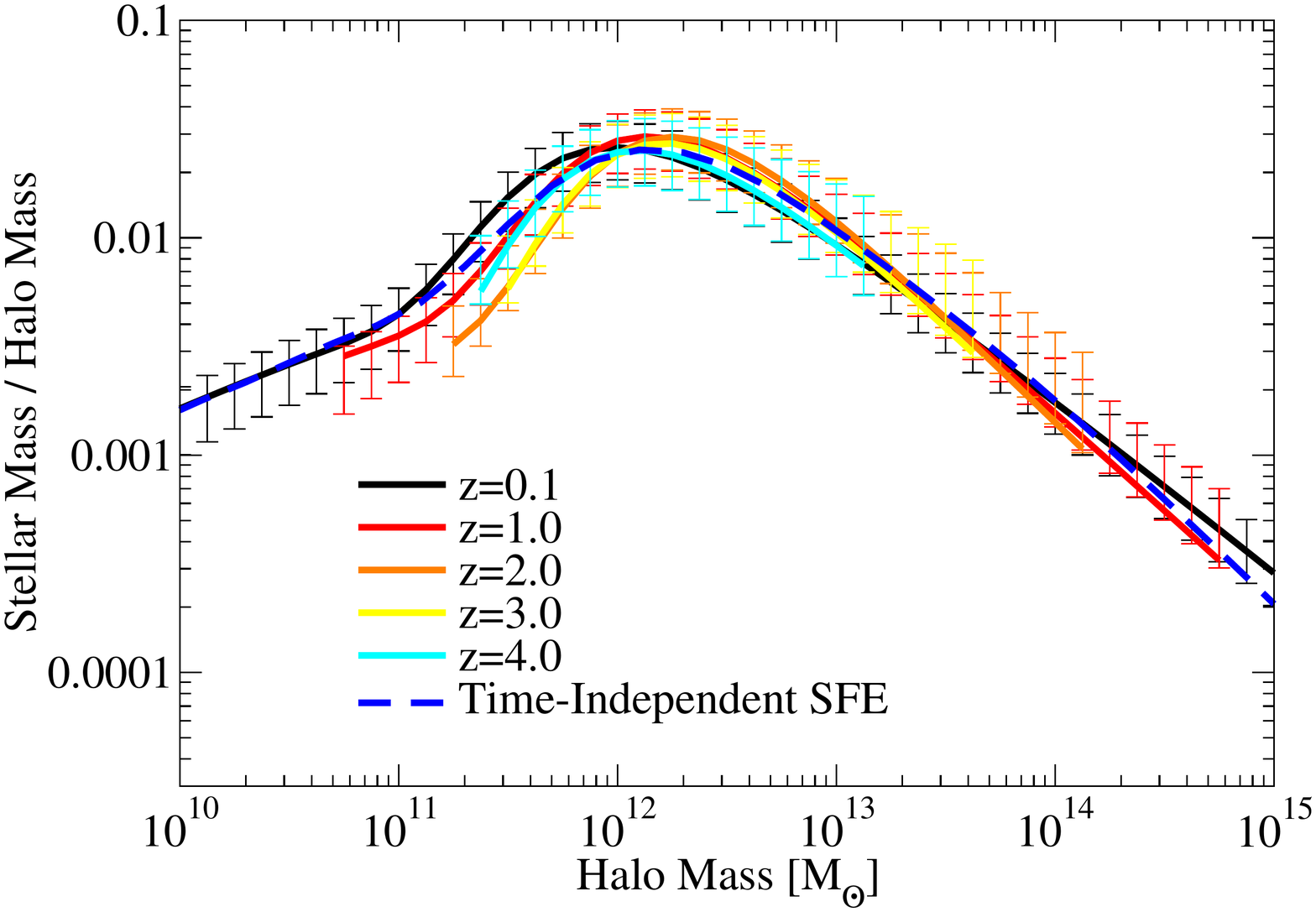,scale=0.33}\epsfig{file=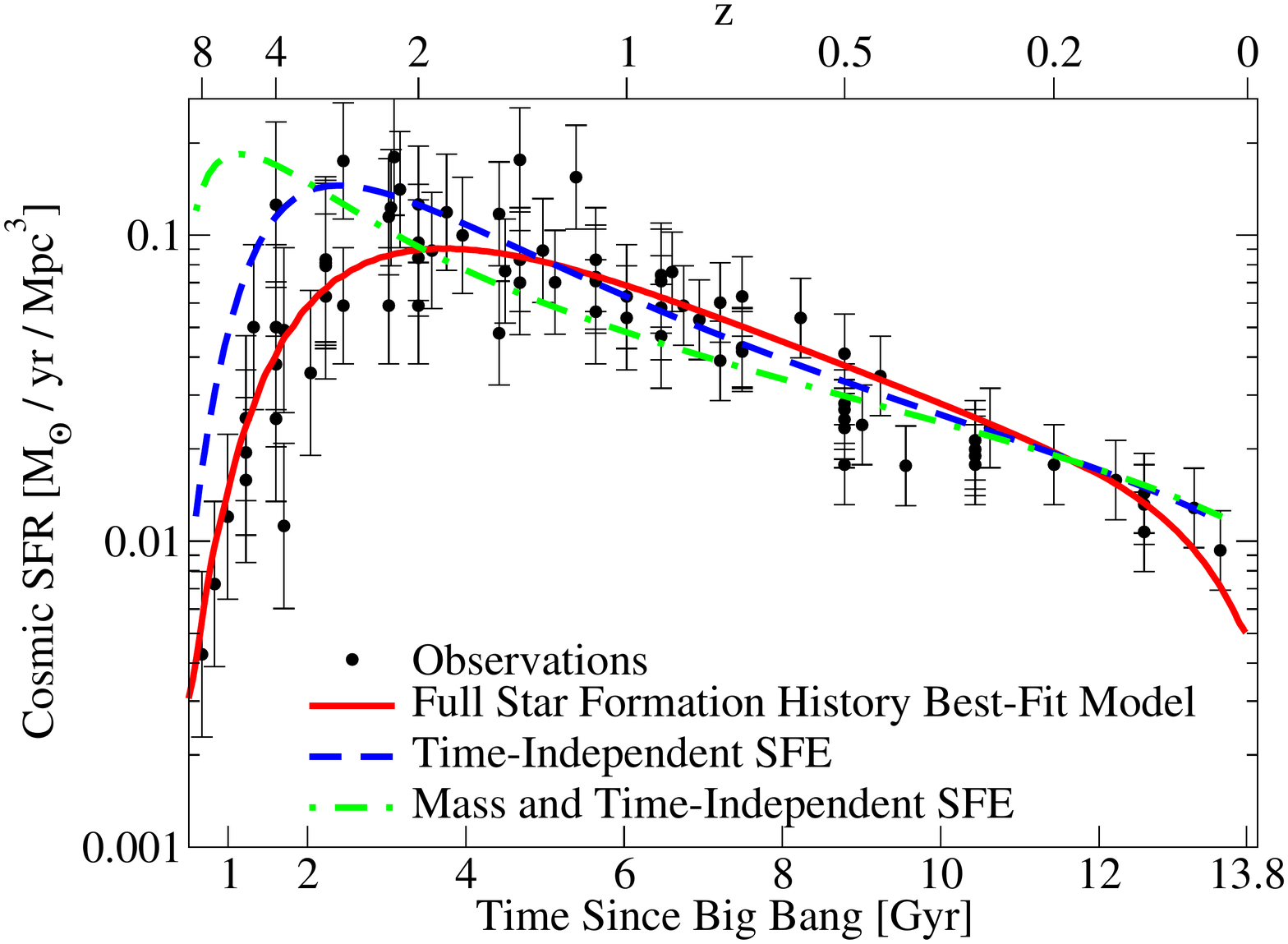,scale=0.33}
\end{tabular}\\
\caption{\textbf{Left} panel: the stellar mass to halo mass ratio at multiple redshifts as derived from observations \citep{BWC12} compared to a model which has a time-independent star formation efficiency (SFE).  Error bars show $1-\sigma$ uncertainties \citep{BWC12}.  A time-independent SFE predicts a roughly time-independent stellar mass to halo mass relationship.  Right: the cosmic star formation rate for a compilation of observations \citep{BWC12} compared to the best-fit model from a star formation history reconstruction technique \citep{BWC12} as well as the time-independent SFE model.  The latter model works surprisingly well up to redshifts of $z\sim 4$.  However, a model which has a constant efficiency (with mass and time) also reproduces the decline in star formation well since $z\sim 2$.}
\label{f:model}
\end{figure*}

\begin{figure*}
\hspace{-12ex}\begin{tabular}{cc}
\epsfig{file=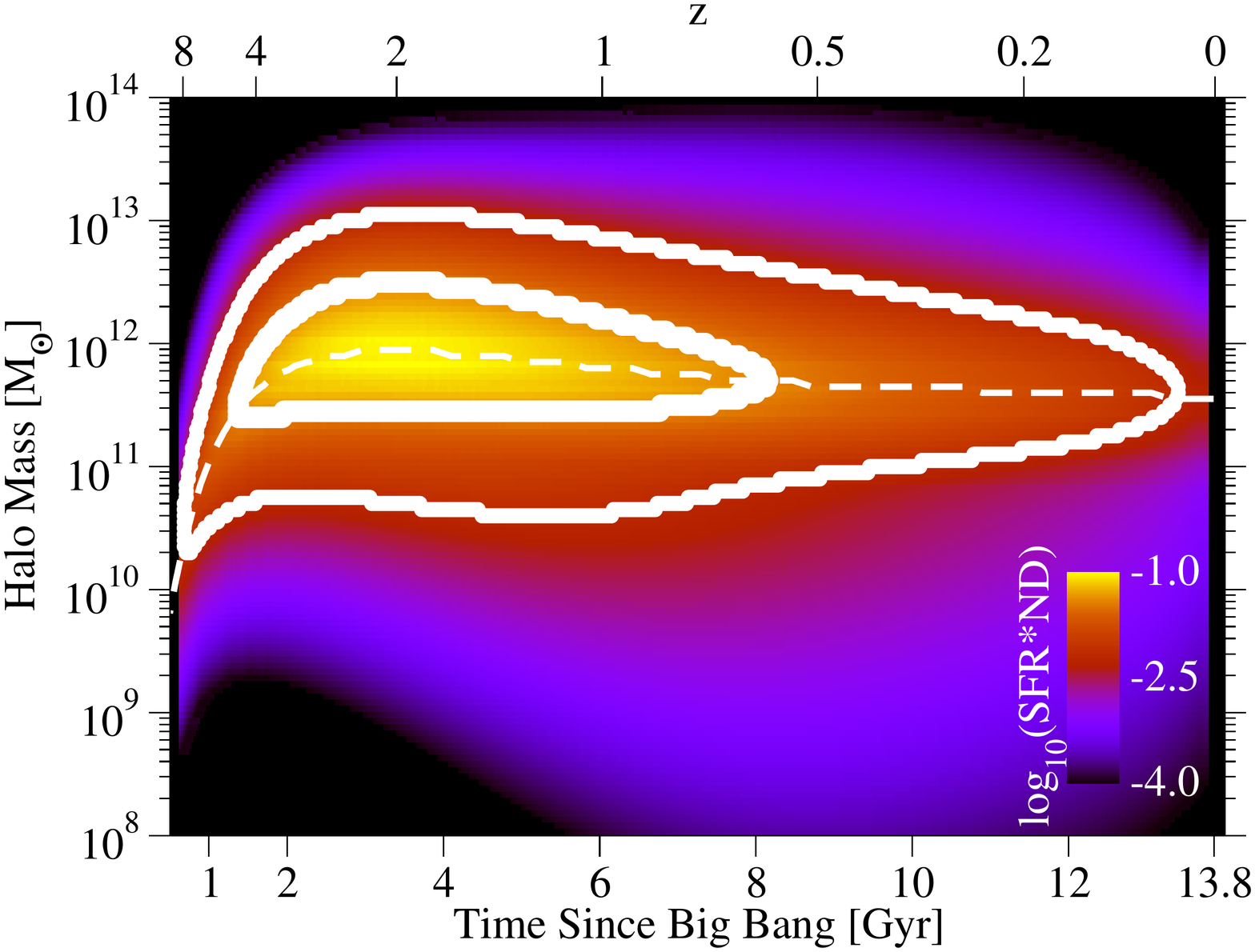,scale=0.33} &\epsfig{file=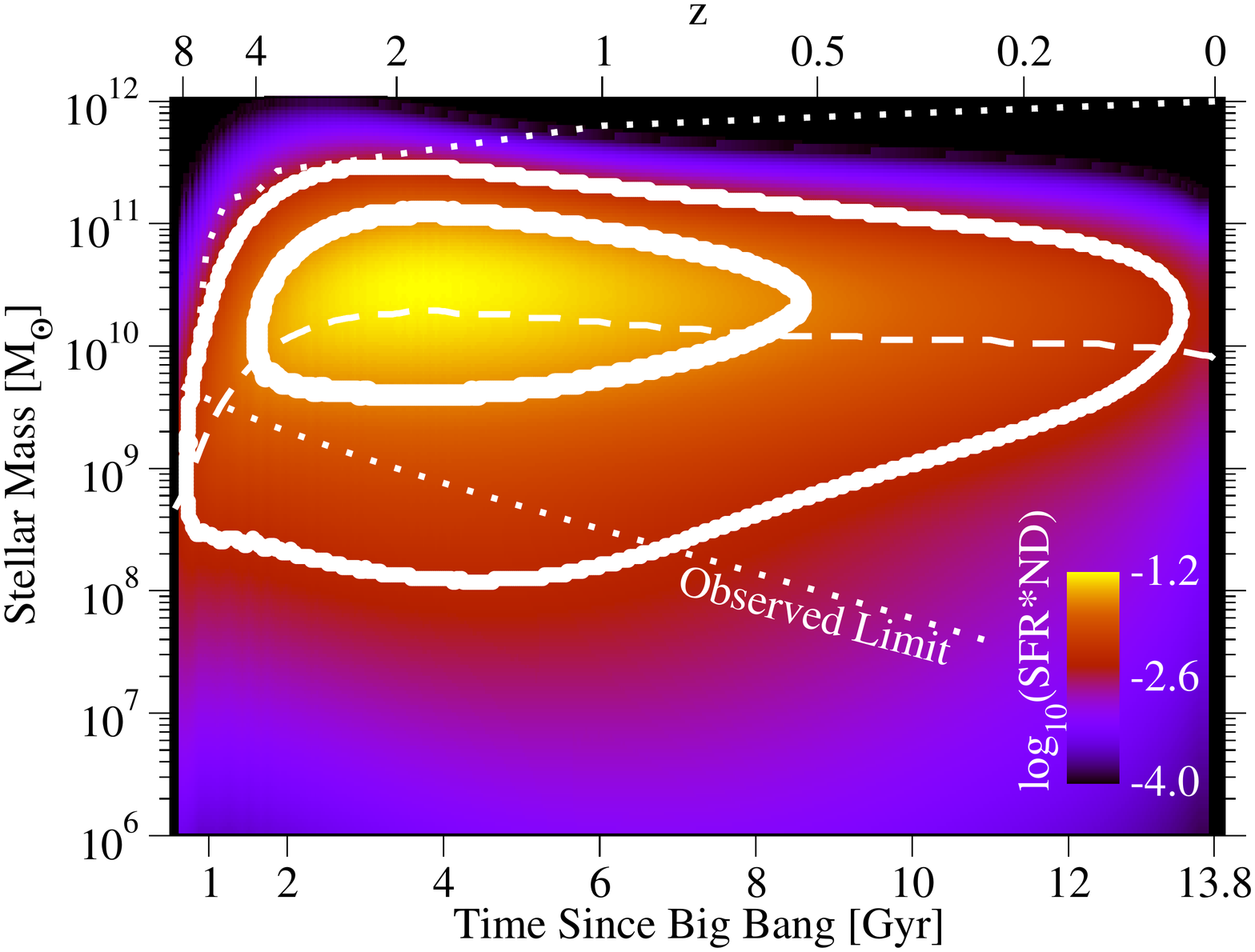,scale=0.33}
\end{tabular}
\caption{\textbf{Left} panel: Star formation rate as a function of halo mass and cosmic time, weighted by the number density of dark matter halos at that time.  Contours show where 50 and 90\% of all stars were formed; dashed line shows the median halo mass for star formation as a function of time.  \textbf{Right} panel: Star formation rate as a function of galaxy stellar mass and time, weighted by the number density of galaxies at that time.  Contours and dashed line are as in top-left panel; dotted line shows current minimum stellar masses reached by observations.}
\label{f:everything}
\end{figure*}

We show the main output of the method in \cite{BWC12}, the average star formation rate (SFR) in dark matter halos as a function of virial mass \citep{mvir_conv} and time, in the top-left panel of Fig.\ \ref{f:bya}.  The SFR depends strongly on time, yet there is also a distinct halo mass threshold, as may be seen by normalizing to the maximum SFR as a function of time (Fig.\ \ref{f:bya}, top-right panel).  To understand the implications for gas physics in halos, it is necessary to consider the baryon accretion rate as well.  
We calculate this as the dark matter halo mass accretion rate \citep{BWC12} times the cosmic baryon fraction; see \cite{vdVoort11} for a comparison with hydrodynamical simulations.

The baryon accretion rate increases with halo mass and lookback time, as shown in Fig.\ \ref{f:bya}, middle-left panel.  This trend combines with trends in the star formation rate to reveal a clear picture of star formation efficiency in halos (Fig.\ \ref{f:bya}, bottom panel).  This efficiency, defined as the star formation rate divided by the baryon accretion rate, shows a prominent maximum near a \massname{} mass of $10^{11.7}\Msun$ (see also Fig.\ \ref{f:integrated}).  Indeed, the star formation efficiency over 90\% of the history of the universe ($z<4$) is strongly dependent on halo mass; by comparison, it has a weak dependence on time.

The peak in the star formation efficiency (SFE) at $10^{11.7}\Msun$ represents observationally-constrained evidence for a 
\massname{} mass for galaxy formation.  This \massname{} mass matches longstanding theoretical predictions
in both its value and in its lack of evolution since $z=3-4$.  The steep efficiency cutoff above the \massname{} mass ($SFE\propto M_h^{-4/3}$, where $M_h$ is halo mass) suggests that a strong physical mechanism prevents incoming gas from reaching galaxies in massive halos.  Besides the effect of hot-mode accretion, this slope coincides with the mass and luminosity scaling for supermassive black holes ($L_{BH} \propto M_{BH}$; scalings for $M_{BH}$ vary from $M_{BH} \propto \sigma^{4} \propto M_h^{4/3}$ to $M_{BH} \propto \sigma^{5} \propto M_h^{5/3}$; \citealt{Ferrarese00,McConnell11}), which may prevent residual cooling flows in massive clusters from forming stars.  Below the \massname{} mass, the efficiency is not a perfect power law; between $M_h \sim 10^{10}$ and $\sim 10^{11.5}\Msun$, the average slope is $SFE\propto M_h^{2/3}$.  This may seem to be consistent with semi-analytic galaxy formation models that use supernova feedback (most commonly scaling as $V_{circ}^2 \propto M_h^{2/3}$; \citealt{Hatton03,Somerville08,Lu12}) to expel most of the gas in low-mass halos.  However, these models often assume that the expelled gas reaccretes onto the halo after a dynamical time \citep{Lu11}; this extra incoming gas would result in a steeper mass-dependence for the SFE at low halo masses.

\subsection{Weak Time Dependence of the Star Formation Efficiency}

The weak time dependence of the star formation efficiency is unexpected given the different environments of $10^{11.7}\Msun$ halos at $z=4$ and $z=0$.  At $z=4$, the background matter density was $\sim$125 times higher, mass accretion rates were $\sim$ 40 times higher, galaxy--galaxy merger rates were $\sim$ 20 times higher, and the UV background from star formation was $\sim$ 500 times more intense than at the present day \citep{BWC12}.  None of these differences significantly influenced average star formation efficiency (unless they conspired to cancel each other out), strongly constraining possible physical mechanisms for star formation in galaxies and halos.

While the time dependence of the SFE is weak, it is not absent.  As seen in Fig.\ \ref{f:bya}, the \massname{} halo mass evolves from a peak of $10^{12} \Msun$ at $z=3$ to $10^{11.5}\Msun$ at $z=0$.  The peak star formation efficiency also evolves around its average value of 0.35, reaching a maximum of 0.55 at $z=0.8$ and a minimum of 0.22 at $z=0$.  However, observational constraints on star formation rates and stellar masses are uncertain at the 0.3 dex level \citep{Behroozi10,BWC12} especially for $z>1$; these are larger than the observed deviations ($\pm$0.2 dex) in the peak star formation efficiency.  The variations in the \massname{} mass are likely more significant; while observational biases can be stellar mass-dependent \citep{BWC12} in a way that changes the location of the peak halo mass, this effect ($<$0.1 dex \citealt{Leauthaud11}) cannot account for the 0.5 dex change from $z=3$ to $z=0$.  Nonetheless, these concerns do not alter the fact that the trends with mass (four decades of variation) in the star formation efficiency are stronger than the trends with time.

One way to eliminate the residual time dependence in the \massname{} mass is to use a different mass definition.  For example, using $M_{200b}$ (i.e., 200 times the background density) would cancel some of the evolution from $z=1$ to $z=0$.  However, this would also raise the mass accretion rate at $z=0$, which would increase evolution in the star formation efficiency's normalization.  Using the maximum circular velocity ($V_{circ}$) or the velocity dispersion ($\sigma$) instead would also lead to more evolution in the SFE (at fixed $V_{circ}$ or $\sigma$): due to the smaller physical dimensions of the universe at early times, both these velocities increase with redshift at fixed virial halo mass.

The nearly-constant \massname{} mass scale is robust to our main assumption that the baryon accretion rate is proportional to the halo mass accretion rate, because this mass scale is already present in the conditional SFR (Fig.\ \ref{f:bya}).  A baryon accretion rate which scales nonlinearly with the dark matter accretion rate would change the width of the most efficient halo mass range, but it would not change the location.  However, as discussed previously, the baryon accretion rate for small halos ($M_h < 10^{12}\Msun$) can differ from the dark matter accretion rate through recooling of ejected gas; the changing virial density threshold can also introduce non-physical evolution in the halo mass which affects the accretion rate \citep{Diemer12}.  Properly accounting for these effects may change the low-mass slope of the star formation efficiency; we will investigate this in future work.

Note that the level of consistency seen in the star formation efficiency is not possible to achieve using other common specific ratios, e.g., the specific star formation rate (SFR to stellar mass ratio; Fig.\ \ref{f:bya}, middle-right panel) or the SFR to halo mass ratio.  The stellar mass --- halo mass ratio (i.e., the integrated formation efficiency) does show somewhat similar features \citep{cw-08,Behroozi10,BWC12,Yang11,Leauthaud11,Moster12,Wang12}; however, the integrated efficiency is several steps removed from the actual physics of star formation.  Galaxy stellar mass is influenced by stellar death, galaxy-galaxy mergers, and ejection of merging stellar mass into the intracluster light \citep{cw-08,BWC12,Moster12}, complicating the interpretation of the integrated efficiency.  Moreover, the shape of the integrated efficiency is influenced by star formation along the entire halo mass accretion history.  Intuitively, the integrated efficiency tends to lag behind changes in the instantaneous star formation efficiency, leading to a peak at a larger halo mass and a gentler fall-off in the high-mass slope, as shown in Fig.\ \ref{f:integrated}.

\subsection{A Time-Independent Model}

Going further, it is interesting to approximate the star formation efficiency for individual halos as completely time-independent.  In this case, the stellar mass formed at a given halo mass is:
\begin{equation}
SM = \int_0^{t_\mathrm{final}} \frac{f_b dM_h}{dt} \frac{dSM}{f_b dM_h} dt = \int_0^{M_{h,\mathrm{final}}} \frac{dSM}{f_b dM_h}f_b dM_h
\label{e:sm}
\end{equation}
(where $SM$ is the stellar mass, $M_h(t)$ is the halo mass accretion history, and $dSM/f_b dM_h$ is the star formation efficiency).  The total stellar mass formed then becomes a function of \textit{only} the final halo mass ($M_{h,\mathrm{final}}$) and not of time.

The specific choice of redshift for the instantaneous star formation rate does not matter greatly, as shown in Fig.\ \ref{f:bya}.  We nonetheless marginalize the instantaneous star formation rate over time; the resulting functional form is shown in Fig.\ \ref{f:integrated}.  Using this as the time-independent efficiency, we calculate the total stars formed as a function of halo mass using Eq.\ \ref{e:sm} and reduce the resulting value by 50\%, corresponding to the stellar population remaining for a 6 Gyr-old starburst \citep{cw-08}.  (For comparison, a 1 Gyr-old starburst would have 60\% of its original stars remaining).  This allows us to calculate the stellar mass to halo mass ratio, as shown in the left panel of Fig.\ \ref{f:model}. Similarly, we may use halo mass accretion rates and number densities along with the same SFE to calculate the cosmic star formation rate (Fig.\ \ref{f:model}, right panel).

The real universe is more complicated, of course; the stellar mass to halo mass relation must evolve weakly to accurately reproduce galaxy number counts \citep{cw-08,moster-09,Moster12,Behroozi10,BWC12,Leauthaud11,Wang12}.  However, integrating a time-independent SFE with respect to halo mass reproduces the $z=0$ stellar mass to halo mass relation to within observational systematics over nearly five decades in halo mass ($10^{10}$ to $10^{15}\Msun$).  Similarly, integrating the SFE times the mass accretion rate and number density of halos gives a precise match to the observed cosmic star formation rate from $z\sim 4$ to the present.

Furthermore, the prediction in time-independent SFE models of fixed stellar mass formed at a given halo mass is not far off from observational constraints at $z=0$ (0.2 dex scatter in stellar mass at fixed halo mass; \citealt{Reddick12}).  The evolution in the median stellar mass to halo mass relation with time, corresponding to an evolution in the star formation efficiency, may then set a lower bound on the scatter in stellar mass at fixed halo mass at the present day.  Conversely, the scatter in stellar mass at fixed halo mass today sets an upper bound on the possible evolution of the median stellar mass to halo mass ratios at earlier times.

When considering the cosmic star formation rate, the time-independent efficiency model may  imply more success
matching galaxy formation physics than is warranted.  In fact, a model with a star formation efficiency of 7\% independent of halo mass or time also matches the decline in cosmic star formation rates (Fig.\ \ref{f:model}, right panel), but would not match the stellar mass to halo mass ratio or galaxy number counts.  For that reason, the decline in the cosmic star formation rate since $z=2$ is more related to declining dark matter accretion rates than changes in how galaxies form stars.  This may explain past successes in reproducing the cosmic SFR with a variety of incompatible physical models \citep[e.g.,][]{Hernquist03,Bouche10,Krumholz12,Dave12} --- the cosmic SFR for $z<2$ alone is a poor discriminant between models.  That said, the rise in the cosmic star formation rate from early times to $z=2$ is much steeper than a mass-independent efficiency model predicts.  Matching this rise is much more closely tied to galaxy formation physics, as it requires an increase in the average star formation efficiency with time.  In the mass-dependent model, this is provided by an increasing number of halos reaching the \massname{} mass.

\subsection{Consequences for When and Where Stars Were Formed}

The star formation efficiency leaves a distinct imprint on the star formation history of the universe: as halos pass through the \massname{} mass ($10^{11.7}\Msun$), they form most of their stars.  Equivalently, most stars were formed in halos between $10^{11.5}\Msun$ and $10^{12.2}\Msun$ (Fig.\ \ref{f:everything}, left panel).   Furthermore, because of the tight correlation between stellar mass and halo mass, most stars formed in galaxies with stellar masses between $10^{9.9}$ and $10^{10.8}M_\odot$ (Fig.\ \ref{f:everything}, right panel).   This same narrow range of halo and stellar masses (which includes the stellar and halo mass of the Milky Way; \citealt{Klypin02, Flynn06, Smith07, Busha11b}) is responsible for most star formation since at least $z=4$, due to the constancy of the star formation efficiency with time.  Given current observational limits (Fig.\ \ref{f:everything}, right panel), surveys have probed a stellar mass and redshift range corresponding to 90\% of the star formation in the Universe.

\section{Conclusions}

\label{s:conclusions}

As we have shown, the ratio of star formation to baryon accretion in galaxies falls off strongly on either side of a \massname{} halo mass and appears to be only weakly correlated with time and environment.  This would suggest a model for galaxy formation in which self-regulation after $z\sim 4$ is nearly perfectly efficient and is controlled by effects which correlate largely with the local gravitational potential: supernova feedback \citep{Dekel86} and possibly metallicity effects \citep{Krumholz12} limit galaxy growth in low-mass halos, and hot mode accretion as well as black hole feedback \citep{Silk98} limit growth in high-mass halos.  Quantitative understanding of how these and other physical feedback effects act to shape observed galaxy formation efficiency will remain a challenge for future research.

\acknowledgments
Support for this work was provided by an HST Theory grant; program number HST-AR-12159.01-A was provided by NASA through a grant from the Space Telescope Science Institute, which is operated by the Association of Universities for Research in Astronomy, Incorporated, under NASA contract NAS5-26555.  This research was also supported in part by the National Science Foundation under Grant No. NSF PHY11-25915, through a grant to KITP during the workshop ``First Galaxies and Faint Dwarfs''.  
We thank Yu Lu, Tom Abel, James Bullock, Louis Strigari, Sandy Faber, Ari Maller, Surhud More, and Joel Primack for insightful discussions during the preparation of this work.\\

\bibliography{master_bib}

\end{document}